# Non-equilibrium VLS-grown stable ST12-Ge thin film on Si substrate: A study on strain-induced band-engineering

S. Mandal[1], B. Nag Chowdhury[1], A. Tiwari[2], S. Kanungo[2], N. Rana[3], A. Banerjee[3], S. Chattopadhyay[1,4,]*

[1]Department of Electronic Science, University of Calcutta, Kolkata, India

[2]Department of Electrical and Electronics Engineering, Birla Institute of Technology and Science-Pilani, Hyderabad, India

[3]Department of Physics, University of Calcutta, Kolkata, India

[4]Center for Research in Nanoscience and Nanotechnology (CRNN), University of Calcutta, Kolkata, India

*Email: scelc@caluniv.ac.in

**Abstract**

The current work describes a novel method of growing thin films of stable crystalline ST12-Ge, a high pressure polymorph of Ge, on Si substrate by a non-equilibrium VLS-technique. The study explores the scheme of band engineering of ST12-Ge by inducing process-stress into it as a function of the growth temperature and film thickness. In the present work, ST12-Ge films are grown at 180 °C – 250 °C to obtain thicknesses of ~4.5-7.5 nm, which possess extremely good thermal stability up to a temperature of ~350 °C. Micro-Raman study shows the stress induced in such ST12-Ge films to be compressive in nature and vary in the range of ~0.5-7.5 GPa. The measured direct band gap is observed to vary within 0.688 eV to 0.711 eV for such stresses, and four indirect band gaps are obtained to be 0.583 eV, 0.614-0.628 eV, 0.622-0.63 eV and 0.623-0.632 eV, accordingly. The corresponding band structures for unstrained and strained ST12-Ge are calculated by performing DFT simulation, which shows that a compressive stress transforms the fundamental band gap at M-Γ valley from 'indirect' to 'direct' one. Henceforth, the possible route of strain induced band engineering in ST12-Ge is explored by analyzing all the transitions in strained and unstrained band structures along with substantiation of the experimental results and theoretical calculations. The investigation shows that unstrained ST12-Ge is a natural n-type semiconductor which transforms into p-type upon incorporation of a compressive stress of ~5

GPa, with the in-plane electron effective mass components at M-Γ band edge to be ~0.09$m_e$. Therefore, such band engineered ST12-Ge exhibits superior mobility along with its thermal stability and compatibility with Si, which can have potential applications to develop high-speed MOS devices for advanced CMOS technology.

**Introduction**

The semiconductor research in recent years is exhibiting a great deal of attention toward germanium (Ge) over again, with the silicon (Si) based mainstream technology being about to reach its scaling limits [1-3]. Alongside exploration of the novel material properties of miniaturized Ge with conventional diamond cubic (DC) crystal structure [1, 4-7], its several pressure-induced allotropes are now being widely investigated [8-12]. Among such allotropes of Ge, a reportedly metastable phase with simple tetragonal lattice structure, *i.e.*, ST12-Ge (also called Ge-III), has drawn significant interest owing to its distinctive optical, electrical and thermoelectric properties [13-19]. The ST12-Ge nano-crystals have been observed to be thermodynamically more stable than DC-Ge in lithium ion battery exhibiting potential energy storage applications [20]. Further, theoretical comparison with hypothetical ST12-Si [21] has led to the prediction of its superconducting behavior at low temperatures [11, 17-18]. Also experimental demonstration shows superconducting property in the mixture of Ge-II and ST12-Ge [22] (although misidentified as BC8-Ge [23]) below ~6 K temperature at a pressure of ~6-10 GPa.

The crystalline structure of ST12 phase of Ge is simple tetragonal exhibiting five-fold and seven-fold rings consisting of 12 atoms per unit cell (*i.e.*, density ~11% higher than DC-Ge) and it belongs to the space group $P4_32_12$ (SG-96) [24]. However, no consensus is developed till date regarding the electronic band structure and band gap of ST12-Ge owing to its metastable behavior posing experimental limitations as well as differences in assumptions and models for theoretical calculations. For instance, the empirical pseudo-potentials with tight-binding model predicts a direct band gap of 1.47 eV [25], whereas less than half of such value (0.7 eV) is obtained from *ab-initio* calculations with plane-wave pseudo-potential under local-density approximation (LDA) [26]. However, a fine sampling of Brillouin zone using the same LDA-based DFT approach results in a fundamental indirect band gap of 0.54 eV and a direct band gap of 0.56 eV [10]. The HSE06 hybrid functional based DFT predicts the indirect and direct band gap to be 0.7 eV and 0.72 eV, respectively [14]. Interestingly, a recent study based on generalized gradient approximation (GGA) shows the band gap of ST12-Ge to be direct with an energy of 0.667 eV, which transforms into indirect band gap with a maximum value of 0.728 eV at a pressure of 2 GPa [19]. However, the experimentally obtained band gap values for bulk ST12-Ge are reported to be 0.53-0.59 eV (indirect) and 0.67-0.74 eV (direct) [19, 14].

Nevertheless, the experimental procedure to synthesize such pressure-induced phase of Ge with reproducibility is not trivial. In fact, depending upon the applied pressure, Ge exhibits numerous structural phase transitions in an irreversible route. For instance, an increasing compressive pressure of ~10-11 GPa on amorphous or DC-Ge transforms it into a metallic phase of β-Sn structure (known as Ge-II) [27, 28], which undergoes a transition to the orthorhombic *Imma* phase at ~75 GPa [29] and to the simple hexagonal (SH) structure at ~85-90 GPa [14, 29]. On application of a compressive pressure of ~ 100 GPa, SH-Ge converts into an orthorhombic phase of Cmca space group and for further compression up to ~160–180 GPa, a hexagonal close-packed (hcp) structure of Ge is obtained [30]. However, the decompression induced phase dynamics of Ge follows a completely different path depending upon entirely different variables. For instance, a rapid decompression (~14 GPa/s) of β-Sn phase to ambient condition transforms it into a metastable BC8 phase, which converts into lonsdaleite (hexagonal diamond) structure in time (~17-24 hrs) [9, 31, 32]. Such fast decompression has also been claimed to yield R8 phase of Ge [33, 34], although no consensus is established due to its reproducibility issues [10]. In contrast, a slow decompression (~1 GPa/day) of β-Sn-Ge leads to the formation of its ST12 phase [28], which makes the synthesis-time prolonged, *e.g.*, 2 days-3 weeks [19]. Interestingly, the β-Sn phase of Ge formed by compressing amorphous Ge transforms into a mixture of BC8 and ST12 phases on decompression [14, 35], whereas β-Sn-Ge obtained by pressurization of crystalline DC-Ge, while being decompressed, yields a mixture of DC-Ge and ST12 [28]. Further, it has been shown that more the shear stress applied to DC-Ge to transform it to β-Sn phase, more will it be converted into the ST12 phase on decompression [36]. Recently, ST12-Ge nanowires are reported to be grown chemically at a moderate temperature of ~290 $^o$C-330 $^o$C [18]. However, ST12 phase being metastable readily transforms into DC-Ge at high temperatures of ~200 $^o$C-300 $^o$C [11].

In this context, the current work reports a non-equilibrium vapour-liquid-solid (VLS) growth of stable ST12-Ge thin film on Si substrate. The film morphology and crystalline quality are investigated by high resolution electron microscopy (FESEM and HR-TEM) and x-ray diffraction (XRD) study. The residual amounts of stress induced in the ST12-films are obtained from Raman spectroscopy and the respective band gap values are estimated from direct and indirect transitions in absorption spectroscopic measurements. The growth temperature is varied to change the film thickness and thereby alter such induced residual stress in ST12-Ge, and accordingly the impact of such stress on the band gap is studied. In order to investigate the origin of such band gap variation with residual stress, the density functional theory (DFT) simulation is performed to comparatively analyze the band structure and the respective valleys associated with optical transitions of unstrained ST12-Ge and that for applying a relevant stress. Through the comparative study of such experimental and simulation results, a comprehensive understanding is developed on the material functionality of ST12-Ge depending on the induced residual stress that leads to the route of its strain induced band engineering.

**Methodological approach**

**A. Sample preparation**

The ST12-Ge thin film is grown on Si substrate in the current work by Au-catalyzed non-equilibrium VLS technique followed by chemical removal of the Au-layer. Prior to such VLS growth, a Si <100> substrate is cleaned according to the standard protocol; for instance, as a first step to remove the organic contaminants by constant ultrasonication in trichloroethylene (TCE), acetone and isopropyl alcohol (IPA) sequentially for 10 min each, and then washing by de-ionized (DI) water; as the second step to remove inorganic and particulate contaminants by heating at 60 °C in a solution of DI water, $H_2O_2$ and HCl in 6:1:1 volume ratio for 10 min and washing by DI water; and as a final step to remove any residual native oxide by dipping into the aqueous HF solution in 1:100 volume ratio for 5 min, followed by drying in gaseous $N_2$ ambient. A ~30 nm Au layer is then deposited on the cleaned Si substrate by thermal evaporation technique at ~$10^{-5}$ mbar pressure. Such Au-coated Si-substrate and Ge-powder (Sigma-Aldrich-327395: ≥99.999%) placed in an alumina boat with an inter-distance of ~1 cm is loaded into a rapid thermal processing (RTP) furnace attached to controlled cooling chiller for the non-equilibrium VLS growth. Then the furnace is heated up to the growth temperature (*e.g.*, 180 °C–250 °C) at a ramp rate of 4 °C/s, kept for 5 min and subsequently cooled down to room temperature at an overall ramp rate of 1-0.05 °C/s as shown in Fig. 1(a) under constant Ar flow of ~140 ml/min. The VLS method with such recipe leads to the formation of a ST12-Ge thin film on Si-substrate underneath the Au-layer. Finally, the Au-layer is removed chemically by sonicating the sample in gold etchant solution (*i.e.*, 1 g of potassium iodide (KI), 0.25 g of iodine ($I_2$) and 10 ml DI water) for 45 s followed by cleaning in DI water [37].

## B. Characterization

The surface morphology along with the film thicknesses are studied by FESEM (Zeiss Auriga), where for imaging the Ge ultra-thin films (thickness ~4.5-7.5 nm) grown on Si-substrate, the scanning is performed at very low accelerating voltage (*e.g.*, 3 kV) to restrict the electron-beam penetration beyond the depth of film thickness and at high magnification (~35 kX) to get finer resolution in the secondary electron (SE) detector. To depict the cross-section for estimating film thickness, a 90º-stub is used and the FESEM stage is rotated to adjust a tilted view at 85º angle (*i.e.*, a grazing angle of 5º) of the top surface of film. The film thicknesses are further verified and more precisely obtained from ellipsometry (Sentech SE 850) study in the visible wavelength range, where the Cauchy relation is used for simulating the ST12-Ge film. The TEM image and selected area electron diffraction (SAED) patterns of the grown film are obtained (JEOL JEM 2100 HR-TEM) at an accelerating voltage of 200 kV and magnification of 800kX. The sample is prepared for such TEM characterization by crushing and grinding the film in an agate mortar to disperse in ethanol followed by ultrasonication and drop-casting on the copper grid. The SAED pattern is analyzed by using CrysTBox [38] simulator to find the diffracting planes of ST12-Ge. The XRD measurement of such film is performed in powder diffraction mode (PANalytical X'Pert Powder) with Cu-k$\alpha_1$ emission line (*i.e.*, the k$\alpha_2$ line is stripped). In order to determine the nature and value of induced residual stress in the grown ST12-Ge thin films, the Raman spectroscopic measurements are performed at room temperature using micro-Raman (Renishaw inVia Raman Microscope) system operating at back-scattering geometry with a 532 nm laser excitation source and a beam spot size of ~1.8 μm$^2$, maintaining the laser power to be 0.05 mW. The absorption behavior of grown Ge-films are obtained by spectroscopic measurement (Perkin Elmer, Lambda 1050) using the integrating sphere in reflection mode, with InGaAs as the IR-detector. The corresponding band gap values are estimated from Tauc plots for direct and indirect transitions.

## C. DFT simulation

The origin of such direct and indirect transitions in the grown ST12-Ge films are investigated from its band structure, computed by performing DFT-based first principle calculations using Atomistic Tool Kit (ATK) and Virtual Nano Lab (VNL) simulation software packages commercially available from Synopsys Quantum-Wise [39]. It is worthy to mention that the DFT simulations are performed in the present work using a linear combination of atomic orbital (LCAO) double-zeta polarized basis set [40], with the density mesh cut-off energy of 100 Hartree and Monkhorst-Pack grid of 6×6×6 for sampling the Brillouin zone (BZ). The ST12-Ge lattices of so-called relaxed (*i.e.*, without stress) and strained (*i.e.*, an applied compressive stress of 5 GPa) configurations are optimized through a limited-memory Broyden-Fletcher-Goldfarb-Shanno (L-BFGS) algorithm [41] with the pressure and force tolerances of 0.0001 eV/Å$^3$ and 0.01 eV/Å, respectively. The geometry optimizations are performed using local density approximation (LDA) [42] method along with Perdew-Zunger (PZ) exchange-correlation functional [43], and finally, the band structures and corresponding electronic properties of ST12-Ge, with and without stress, are calculated using the generalized gradient approximation (GGA) [44] method together with the Purdew-Burke-Ernzerhof (PBE) functional [45].

## Results and Discussion

### A. Growth mechanism

Ge thin films are usually deposited by RF magnetron sputtering [46-48], molecular beam epitaxy [49], thermal evaporation [50], electron-beam evaporation [51, 52], and chemical vapour deposition [53] techniques, whereas Ge nanowires are often grown in a comparatively more cost-effective VLS method [54]. In the current work, however, the VLS method is exploited for growing a thin film of ST12-Ge on Si-substrate, the temperature profile being depicted in Fig. 1(a). The VLS growth of thin films and nanostructures are generally attributed to the formation of a eutectic melt of metal catalyst (*e.g.*, Au) with a source material (*e.g.*, Ge powder) in vapour form providing a smooth interface with the solid substrate (*e.g.*, Si) at an elevated temperature. Such equilibrium of all the three phases leads to precipitation of the source material at supersaturation condition due to Gibbs' free energy minimization resulting in its crystalline growth on the substrate [55-60]. However, the lattice mismatch and thermo-elastic mismatch between the film and substrate materials may incorporate substrate-induced and process-induced strains into the grown film, respectively [59, 61]. Such phenomenon of stress incorporation is utilized in the present work to develop the pressure-induced ST12-phase of Ge film on Si. At room temperature, DC-phases of Ge and Si have a lattice mismatch of ~4%, which increases with temperature since Ge has a higher thermal expansion coefficient (~5.9×10$^{-6}$/$^o$C) than that of Si (~2.6×10$^{-6}$/$^o$C). Thus, during the VLS growth at higher temperatures (*e.g.*, 180 $^o$C-250 $^o$C), such increased lattice-mismatch associated with thermal-mismatch incorporates stress into the Ge film. Since Si-substrate is elastically much stable than Ge thin film, the lattice of Ge (5.658 Å) gets compressed to accommodate a lattice constant almost equal to that of Si (5.431 Å) to minimize the formation energy. However, such induced pressure is relaxed when the system is cooled down. At this point, it is important to note that an irreversible but slow decompression of strained (compressive) Ge leads to its transformation into the ST12-phase. For such purpose, the entire VLS growth of Ge film is carried out in the present work inside a RTP-furnace attached to controlled cooling chiller, which is capable of providing very fast temperature ramp rate. After studying the film growth for various cooling rates, an optimized cooling rate of 1-0.1 °C/s for 2 mins followed by a constant rate of 0.05 °C/s is maintained for all the samples for different growth temperatures in the current work. Such faster cooling rate of the RTP-furnace (compared to natural Newton's cooling) leads to a non-equilibrium, and thus evidently an irreversible, thermal decompression of the Ge thin film developed on Si substrate. The decompression rate is estimated from such cooling rate and the corresponding temperature dependent lattice mismatch between phases of Ge and Si to be ~1 MPa/s. Such slow but non-equilibrium decompression of Ge results in the formation of its ST12-phase in the present work.

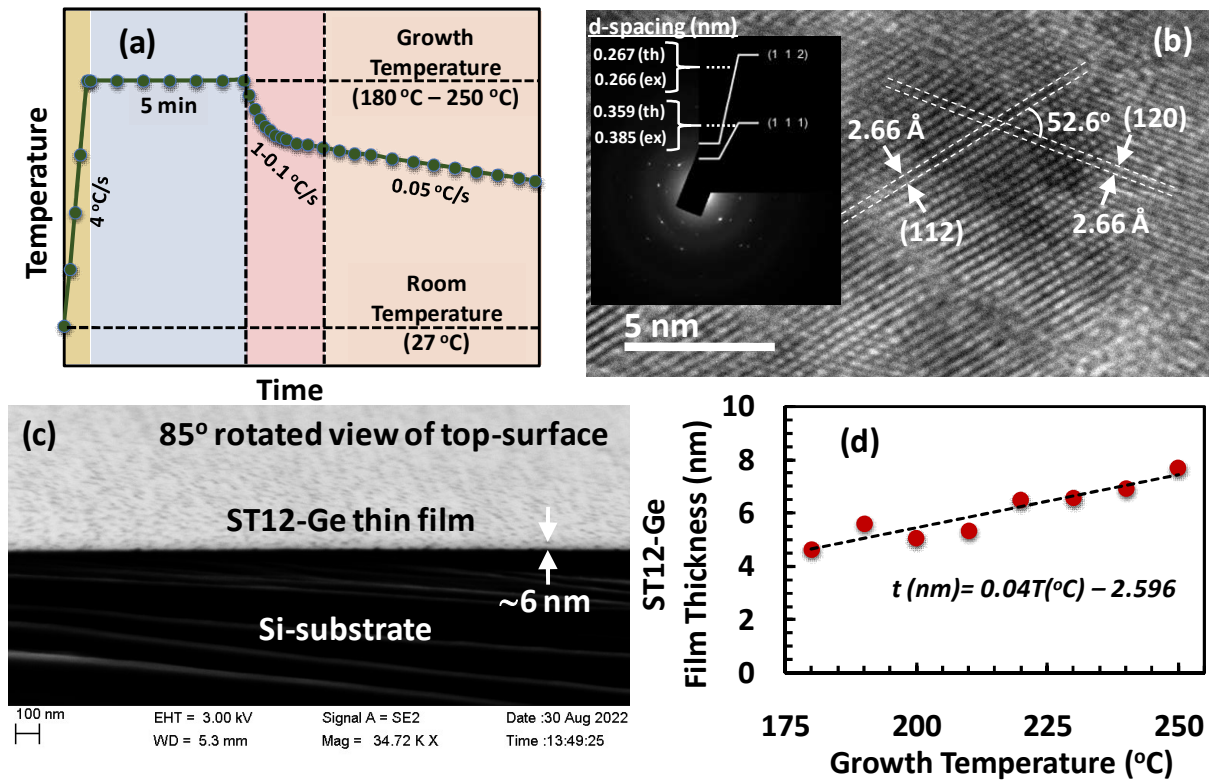

Fig. 1(a). The temperature profile for VLS growth of ST12-Ge film on Si-substrate, the 'pink' region indicating non-equilibrium decompression; (b) HRTEM image of the grown film indicating (112) and (120)-planes of ST12-Ge, with the SAED pattern and simulated planes as inset; (c) FESEM image of the cross-section of ST12-Ge/Si at a grazing angle of 5$^o$ (*i.e.*, 85$^o$ rotated from top-view) and EHT=3 kV; (d) plot of film thickness with growth temperature showing a linear fit.

**B. Crystal structure and film quality**

The HRTEM image of the grown Ge film is shown in Fig. 1(b), where the SAED spots along with the simulated planes (using CrysTBox) are given in the inset. The TEM and electron diffraction analysis confirm the growth of (112)-plane of ST12-Ge, with the d-spacing obtained to be ~2.66 Å. It is worthy to note that the same lattice positions of the atoms are also arranged in the (120)-plane of ST12 at an angle of ~52.6$^o$ with its (112)-plane, as shown in Fig. 1(b). Therefore, simulating the same diffraction spots considering (120)-plane, the d-spacing is calculated and obtained also to be ~2.66 Å, which is also consistent with the value estimated from the TEM image. The results are in very good agreement with the theoretical values of such d-spacings, *i.e.*, 2.67 Å and 2.65 Å, respectively for (112) and (120)-planes, as well as with the previous experiemntal reports [13]. Further, a lesser number of SAED spots indicate (111)-plane with d-spacing of 3.85 Å, which is comparable with the theoretical value of 3.59 Å for bulk ST12-Ge. Such small deviations in d-spacing values are reported to appear in thin films and nanocrystals, however, cannot change the internal symmetry of the crystalline structure [13].

FESEM image of the grown Ge film is illustrated in Fig. 1(c), which depicts the cross-section of ST12-Ge/Si at a grazing angle of 5$^o$ (*i.e.*, 85$^o$ from top-view). It is apparent from the figure that a continuous film is grown in the present recipe of VLS-method and the film thickness obtained to be ~6 nm. In fact, for the samples grown at several temperatures (*e.g.*, 180 $^o$C-250 $^o$C), the thickness is observed to vary within 4-8 nm. Such thicknesses are further calculated from ellipsometric study considering the Cauchy relation for ST12-Ge, and plotted with growth temperature in Fig. 1(d). The study suggests that the film thickness of ST12-Ge grown on Si in VLS method followed by chemical removal of the Au-layer exhibits an almost linear dependence on the growth temperature. It can further be predicted empirically from such linear relation that no such film would grow at a temperature below ~65 $^o$C (*i.e.*, film thickness ~ 0 nm in Fig. 1(d)).

## C. XRD and stability of ST12-phase

As can be observed from the plots of Fig. 2, all the ST12-Ge films grown in the present VLS method exhibit a sharp XRD peak at 2θ=33.12° indicating its (112)-crystal plane in reference to JCPDS No. 72–1089. It is imperative to mention that, although bulk ST12-Ge shows several XRD peaks with strongest reflection from such (112)-plane [14, 19], ST12-Ge nanowires are reported to observe such single peak due to the crystallographic orientation of their axes to be along (112)-plane [17]. Further, the TEM analysis and SAED patterns (Fig. 1(b)) confirm the strongest transmission/diffraction from its (112)-plane. Therefore, the present XRD results indicate the VLS growth of ST12-Ge film along (112)-plane vertically on Si substrate.

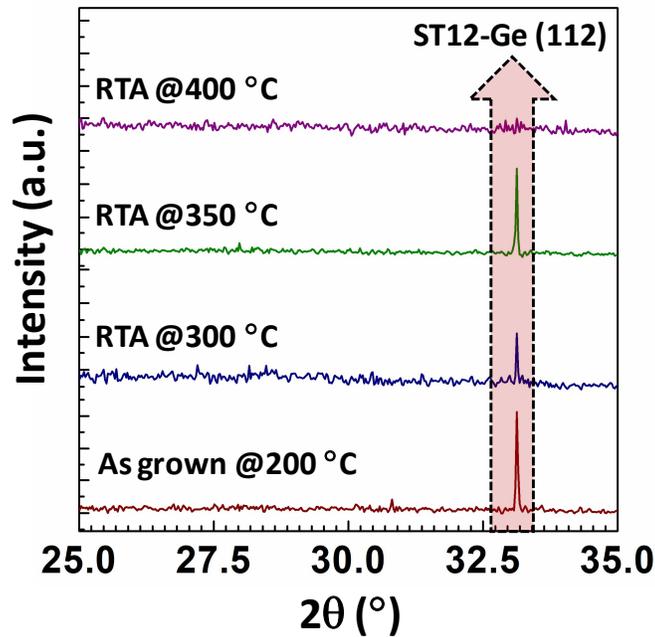

Fig. 2. XRD patterns of ST12-Ge film grown at 200 °C and after rapid thermal annealing at 300 °C, 350 °C and 400 °C.

It is further worthy to mention that the bulk ST12-Ge has been reported to loss its phase stability completely through a transformation into DC-Ge for annealing at 255 °C [14]. In order to investigate the thermal stability of such phase in the present samples, the as-grown films are annealed under rapid thermal process (RTA) at 300 °C, 350 °C and 400 °C, and the relevant XRD patterns are plotted in Fig. 2. It is evident from the plots of Fig. 2 that the ST12-phase of Ge film grown on Si by the current VLS method remains stable at least up to 350 °C, while it is observed to lose its phase stability for annealing at 400 °C. Thus, the thermal stability of VLS-grown ST12-Ge film is significantly improved compared to the previously reported bulk samples fabricated by the conventional compression-decompression techniques. A close inspection shows very small peaks to appear at 2θ~27.5°-28° for annealing the samples, which may be attributed to the transformation of ST12 into DC-Ge of (111) plane, which is also consistent with the previously published reports [14].

### D. Raman scattering and induced residual stress

The room temperature Raman spectra of ST12-Ge films grown at different temperatures are plotted in Fig. 3(a), where the respective peaks associated with the relevant Raman active phonon modes [17, 19] are indexed. It is imperative to mention that the Raman peak observed at ~302 cm$^{-1}$ is due to Si substrate (verified by standard Si wafer without Ge) and does not indicate any presence of DC-Ge. In the current work all the grown ST12-Ge films exhibit $B_1$, $B_2$ and E optical vibrational modes representing the anti-symmetric about principal axis of symmetry, totally anti-symmetric and doubly degenerate phonon modes, respectively [17]. It is apparent from Fig. 3(a) that the Raman peak positions are right shifted from the defined phonon modes which are attributed to compressive stress induced into the ST12-Ge films. Such stress is originated from the thermo-elastic and lattice mismatch between Ge and Si and incorporated during the VLS-growth as discussed earlier. The induced stress values in the grown films are plotted with film thickness in Fig. 3(b).

At this point it is worthy to mention that the net stress ($\sigma$) induced in a thin film (of thickness $t_f$ and Young's modulus $Y_f$) grown on a substrate (of thickness $t_s$ and Young's modulus $Y_s$) due to a temperature change ($\Delta T$) under the condition $t_f/t_s \ll 1$ is given by Roark's Formula [62] as $\sigma = \sigma_T\left(-1 + 4Y_{fs}(t_f/t_s) + 3Y_{fs}(t_f/t_s)^2\right) + \sigma_M Y_{fs}\left(1 - 2(2Y_{fs} - 1)(t_f/t_s)\right)$, where $Y_{fs} = (Y_f/Y_s)$, $\sigma_T = Y_f \Delta\alpha \Delta T$ is associated with the thermal strain components, $\Delta\alpha$ being the mismatch in thermal expansion coefficient, and $\sigma_M$ is the stress related to bending moments. The induced stress in the current VLS-grown ST12-Ge films on Si substrate are observed to be compressive in nature and vary in the range of ~0.5-7.5 GPa for film thicknesses ~4.5-7.5 nm, as shown in Fig. 3(b). It is apparent from Fig. 3(b) that for lower film thicknesses (<6.5 nm), the induced stress exhibits an almost parabolic variation indicating the thermal strain to be more significant than that of bending moment. However, the bending moment is observed to play the key role for higher film thicknesses (≥6.5 nm) and almost a linear decrement of stress is found, which is expected since a larger film thickness suffers lesser strain due to crystal stability.

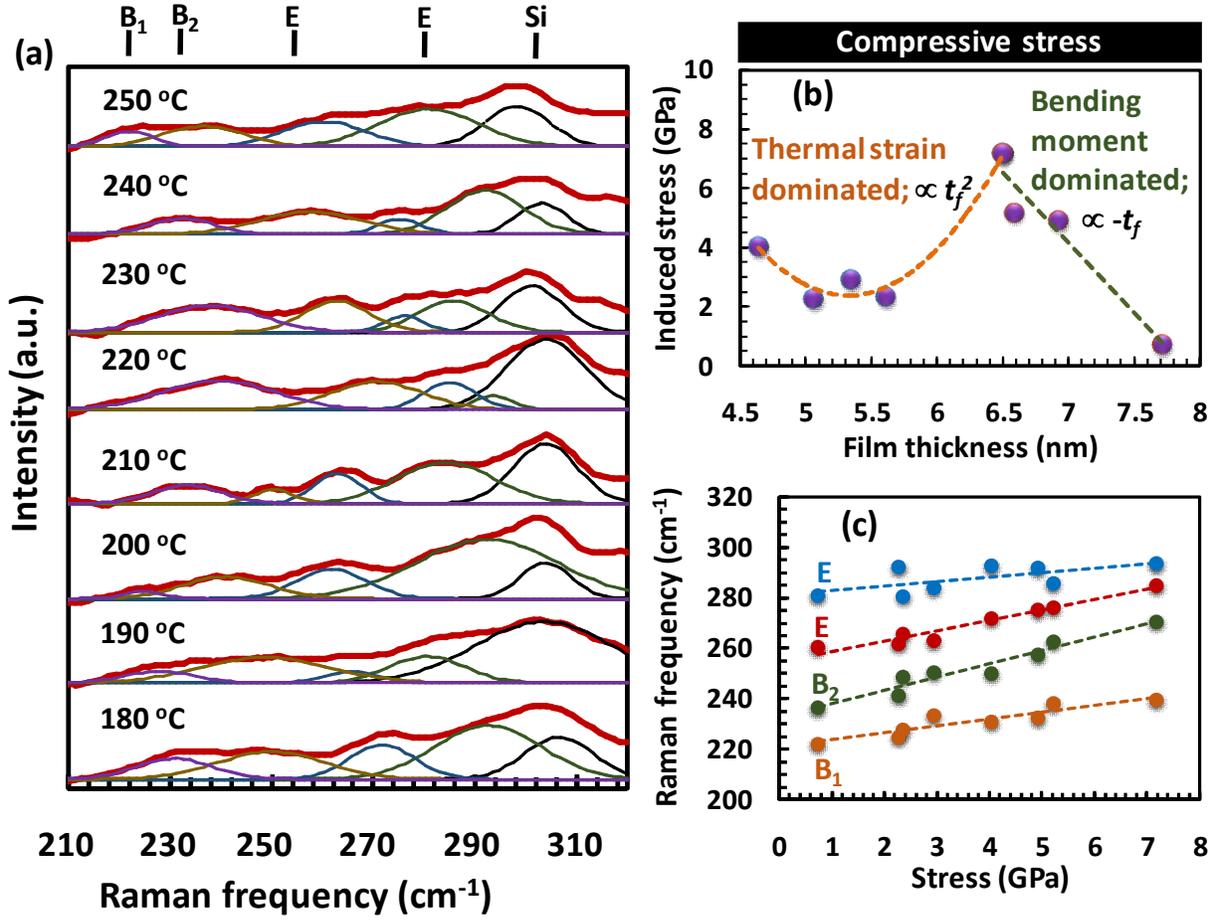

Fig. 3 (a). Room temperature Raman spectra of ST12-Ge films grown at different temperatures (180 °C – 250 °C) showing the deconvoluted peaks along with the peaks associated with unstrained Raman active phonon modes; (b) the plot of induced compressive stress in the grown ST12-Ge films with film thickness distinguishing the thermal strain dominated and bending moment dominated regions; (c) the variation of Raman frequency with induced compressive stress values for the observed modes.

The stress values in ST12-Ge films are estimated in the current work from corresponding Raman peak shifts by solving the secular equation considering relevant Raman tensor components [63] for its (112)-plane (since growth along such plane is confirmed from TEM and XRD). It is imperative to note that, for back scattering from (112)-plane, only the Z-component of Raman tensor ($R_z$), indicative of transverse optical (TO) phonon modes, can be observed in experiment. Therefore, the solution of secular equation [63] in the present case of ST12-Ge films grown on Si substrate, considering biaxial stress (since two in-plane directions are symmetric), leads to $\omega_j - \omega_{0j} = G(\omega_{0j})\sigma$; where, $\omega_j$ and $\omega_{0j}$ are strained and unstrained Raman frequencies, $\sigma$ is the value induced stress and $G(\omega_{0j})$ depends on the phonon deformation potentials and elastic compliance tensor [63] of the material (*i.e.*, ST12-Ge). It is worthy to mention that an almost linear dependence of $G(\omega_{0j})$ on the unstrained phonon modes $\omega_{0j}$ indicates negligible anharmonicity in atomic vibrations.

Table. I. The summary of Raman active phonon modes for unstrained ST12-Ge thin film grown in the present VLS method and comparison with other theoretical and experimental reports.

| Raman active phonon mode | Refs. | Theoretical value (cm$^{-1}$) | Extracted experimental value (cm$^{-1}$) | Experimental value of present work (cm$^{-1}$) |
|---|---|---|---|---|
| B1 | [9] [14] [17] [19] | 227.3 213.8 | 221.9 215.8 225 (±4) 213.0 | 221.1 |
| B2 | [9] [14] [17] [19] | 228.4 229.3 | 241.8 231.9 232 (±2) 230.0 | 232.7 |
| E | [9] [14] [17] [19] | 255.8 254.5 253.3 | 256.9 247.8 261 (±4) 246.9 | 254.7 |
| E | [9] [14] [17] [19] | 282.7 272.8, 281.0 | 270.4 276.8 275 (±5) 273.8 | 281.3 |

In the current work, the values of $G(\omega_{0j})$ for ST12-Ge are extracted from the previously reported experimental results of [9], which show a very good linear fit of $\omega_j - \sigma$ plot. Using such values of $G(\omega_{0j})$ in the frequency shift, $\omega_j - \omega_{0j} = G(\omega_{0j})\sigma$, the values of stress, induced in the present VLS-grown ST12-Ge films, are calculated for the observed phonon modes. The experimentally obtained Raman frequencies and the induced compressive stress values are thereby plotted in Fig. 3(c) for each of such observed modes. The unstrained phonon modes for ST12-Ge are then estimated from the intercepts of linear fit of $\omega_j - \sigma$ plot for each observed mode, and compared with the previous theoretical and experimental reports by summarizing in Table. I.

### D. Absorption, band structure and impact of stress

The IR-absorption spectra of ST12-Ge films grown at various temperatures (180 °C – 250 °C) are plotted in Fig. 4, which show that each of the grown samples exhibits four absorption-peaks and one absorption-saturation. The corresponding optical band gaps are calculated from the standard Tauc plots of $(ah\nu)^2$ and $(ah\nu)^{1/2}$ with $h\nu$ for direct and indirect transitions, respectively; where $a$ is the absorption coefficient, $\nu$ be the frequency of incident light and $h$ Planck's constant.

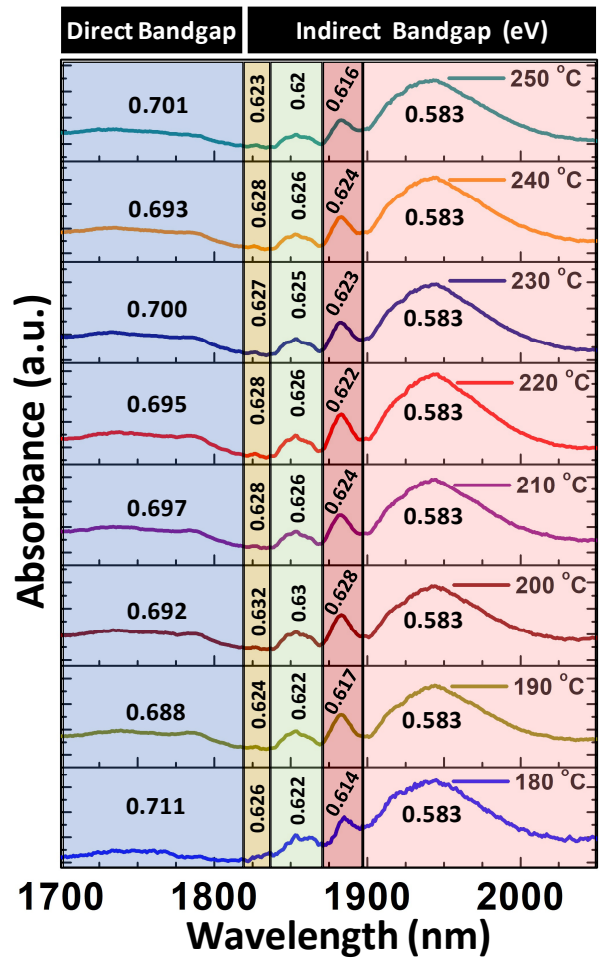

Fig. 4. The plots of IR-absorption spectra of ST12-Ge films grown at 180 °C – 250 °C indicating the direct and indirect band gaps, calculated from Tauc plots of $(ah\nu)^2$ and $(ah\nu)^{1/2}$ against $h\nu$, respectively.

As indicated in Fig. 4, the optical band gaps are observed to be ~0.583 eV for indirect transition (right most peak) and it varies in the range of 0.688-0.711 eV for direct transition (left most peak) in the grown films, which are in very good agreement with the previously reported experimental results [14, 19]. However, it is apparent from Fig. 4 that the present VLS-grown ST12-Ge films exhibit three more optical transitions that are associated with very close indirect band gaps in the range of 0.614-0.628 eV, 0.622-0.63 eV and 0.623-0.632 eV, which are not reported in the previously published works.

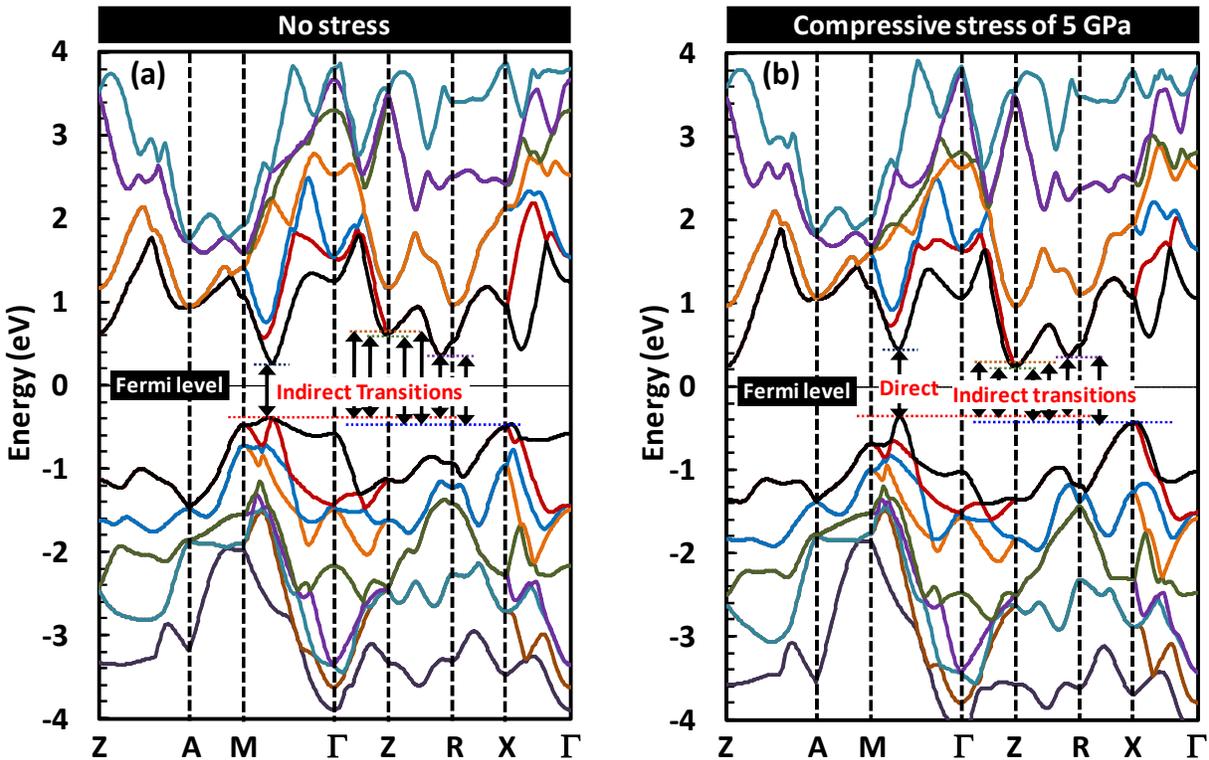

Fig. 5. DFT simulated band structure of ST12-Ge under GGA considering PBE-functional: (a) for no stress applied; and (b) for a compressive stress of 5 GPa.

In order to investigate the origin of such optical transitions, the band structures of ST12-Ge are estimated by DFT-based first principle calculations using GGA along with PBE functional for no-stress condition and a compressive stress of 5 GPa. At this point it is worthy to mention that the calculated lattice parameters of the geometry-optimized so-called 'relaxed' (*i.e.*, no stress) ST12-Ge crystal are found to be a=b=5.89 Å, and c=6.88 Å, which are in excellent agreement with the previously reported experimental and theoretical values **[14, 19]**, and justifies the reliability of the entire theoretical framework employed in the present study. Further, the compressive stress configurations in ST12-Ge films are considered to be in-plane biaxial (*i.e.*, along 'a' and 'b' lattice vectors), preserving the hexagonal symmetry of the crystal. The resultant band structures of ST12-Ge, with and without stress, are plotted in Fig. 5(a) and (b), respectively, and the possible direct and indirect optical transitions are indicated accordingly.

As indicated in Fig. 5(a), the fundamental band gap of 'relaxed' ST12-Ge at M-Γ valley is 'indirect', with a value of ~0.66 eV, since its corresponding conduction and valence band edges appear at (0.34,0.34,0) and (0.36,0.36,0) *k*-points, respectively. Such results are in complete agreement with the previously reported DFT calculations, *i.e.*, 0.667 eV (GGA-PBE) **[19]** and 0.7 eV (HSE06) **[14]**. The other possible indirect transitions from two valence band maxima (VBM) at M-Γ and X to the conduction band minima (CBM) at R-Z-point and two quasi-degenerate states at Z-point are also shown in Fig. 5(a). Interestingly, it is observed from the plots of Fig. 5(b) that the fundamental band gap of ST12-Ge at M-Γ valley transforms from 'indirect' to 'direct' for 'compressive stress', with the VBM shifting to (0.34,0.34,0) *k*-point. The present GGA-PBE calculation yields such direct band gap to be 0.8 eV for 5 GPa compressive stress, highly consistent with the present experimental result of 0.7 eV for 5.2 GPa (*i.e.*, the film grown at 230 °C). The small overestimation of ~0.1 eV may be attributed to the thermal effects that are not considered in the DFT calculation **[14]**. Further, as can be seen from Fig. 5(b), the indirect band gaps between M-Γ and the quasi-degenerate Z are 0.594 eV and 0.605 eV leading to the optical transitions, which are possibly convoluted in the absorption spectra (Fig. 4), exhibiting a broadened peak around the band gap of 0.583 eV. However, the indirect transitions from X to the two quasi-degenerate states at Z with theoretical energy gaps of 0.666 eV and 0.677 eV are observed as two corresponding peaks in absorption spectra, indicating the experimental values of such band gaps to be 0.623 eV and 0.625 eV, respectively. The experimental indirect band gap of 0.627 eV may be attributed to the optical transition from M-Γ to R-Z, where the theoretical calculation slightly overestimates it by ~0.07 eV. The X to R-Z transition with a gap almost equal to that of direct transition at M-Γ is possibly overlapped in the absorption spectra and therefore not observed distinctly. It is further interesting to note from the present DFT calculation that the unstrained ST12-Ge is naturally n-type with its Fermi level located ~0.06 eV above the intrinsic level. However, a 5 GPa compressive stress converts it into a p-type semiconductor with the Fermi level at ~0.05 eV below its intrinsic level. At such condition, the in-plane electron effective mass components at M-Γ CBM are obtained to be $0.093m_e$, which indicates a significantly higher mobility than Si and therefore suggests its possible application in high speed MOSFET devices.

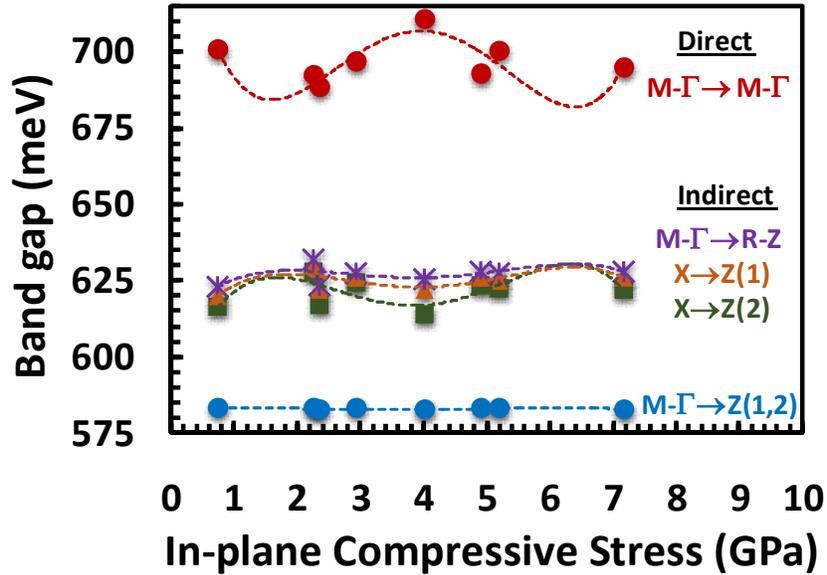

Fig. 6. The variation of experimentally obtained band gap (direct and indirect) values with in-plane compressive stress in the ST12-Ge films indicating the associated valley points for optical transitions. A fourth order polynomial of is fitted for all the transitions.

It is apparent from Fig. 5(a) and (b) that the direct band gap increases for a compressive stress of 5 GPa, whereas the indirect band gaps are observed to decrease. Such opposite nature of band gap change is attributed to the fact that the energy of M-Γ valley (*i.e.*, $k_z = 0$), that represents in-plane biaxially symmetric electron transport, increases for in-plane biaxial compression due to enhanced atomic wavefunction overlap; however, such in-plane compression results in a corresponding out-of-plane tensile stress (due to positive Poisson's ratio of ~0.19 [19]) that reduces the energy of valleys associated with Z-point (*i.e.*, $k_x = k_y = 0$). Such opposite trend can evidently be observed in the variation of experimental band gap (direct and indirect) values with in-plane compressive stress in ST12-Ge films, as plotted in Fig. 6, where the corresponding valley points for optical transitions are indicated as per DFT calculations. Interestingly, the indirect band gap associated with M-Γ→Z transitions do not exhibit notable change which is attributed to the opposite effects of in-plane biaxial compression on M-Γ and out-of-plane tensile stress on Z-point. At this point it is worthy to mention that, since ST12-phase of Ge is itself a result of decompression from its β-Sn phase and the VLS growth process induces a residual compressive stress into it, the countering effects of such inherent decompression and induced compression leads to a maximum in the direct band gap and minimum in indirect band gap associated with in-plane compression and out-of-plane tensile stress, respectively (as apparent from Fig. 6). The nature of experimentally obtained band gap values with increasing stress is fitted with polynomial curve, which shows a fourth order variation for all the transitions, and thus predicts the unstrained band gap values of ST12-Ge films to be 0.756 eV for direct transition and 0.610 eV, 0.601 eV, 0.586 eV and 0.582 eV for M-Γ→R-Z, X→Z(1), X→Z(2) and M-Γ→Z indirect transitions, respectively.

**Conclusion**

In summary, the current work explores a method of growing stable crystalline ST12-Ge thin film on Si substrate by a non-equilibrium VLS technique and its band engineering by inducing stress depending on growth temperature and film thickness. The ST12-Ge films are observed to grow at 180 °C – 250 °C along (112)-crystal direction and exhibit very good thermal stability up to a temperature of ~350 °C. Thicknesses of the grown films are obtained to be ~4.5-7.5 nm and show a linear variation with growth temperature. The stress induced in such ST12-Ge films are measured from Raman spectroscopic study and found to be compressive in nature varying in the range of ~0.5-7.5 GPa. The origin of such stress is attributed to the thermo-elastic and lattice mismatch between Ge and Si; and observed to be thermal strain dominated up to a film thickness of ~6.5 nm and then conquered by the residual bending moments. The absorption spectroscopy shows five distinct optical transitions providing a direct band gap of 0.688-0.711 eV, and indirect band gaps of 0.583 eV, 0.614-0.628 eV, 0.622-0.63 eV and 0.623-0.632 eV. The origin of such band gaps is investigated by calculating the band structure of ST12-Ge from DFT simulation under GGA using PBE functional. Its fundamental band gap at M-Γ valley is observed to transform from 'indirect' (without stress) to 'direct' for compressive stress. The other indirect band gaps remain 'indirect' and are attributed to the optical transitions from M-Γ and X to R-Z and the quasi-degenerate Z points. However, the trend of band gap change with stress for direct and indirect transitions is observed to be opposite in nature due to in-plane compression and out-of-plane tensile stress. Such band gap changes with induced stress exhibit an extremum due to the countering impact of inherent decompression in ST12-phase and the thermal compression during its growth. Further, the DFT calculations reveal that unstrained ST12-Ge is natural n-type semiconductor whereas converts into p-type for a compressive stress of ~5 GPa. Most importantly, the in-plane electron effective mass components of ST12-Ge at M-Γ band edge is obtained to be ~0.09$m_e$ that suggests its significantly higher mobility than Si. Therefore, such cost-effective VLS growth of ST12-Ge film with its superior mobility, thermal stability and compatibility with Si can be utilized for significant performance improvement in advanced CMOS technology.

## Acknowledgement

S. Mandal would like to acknowledge the Council of Scientific and Industrial Research (CSIR: 09/028(1103)/2019-EMR-I) for providing fellowship. The authors would like to acknowledge DST PURSE, Center of Excellence (COE) for Systems Biology and Biomedical Engineering, and Centre for Research in Nanoscience and Nanotechnology (CRNN) for providing infrastructural support to conduct this work.